\documentclass{emulateapj}
\usepackage{apjfonts}

\begin{document}

\title{Very high efficiency photospheric emission in
long duration $\gamma$-ray bursts}
\shorttitle{GRB Photospheres}

\author{Davide Lazzati\altaffilmark{1,2}, Brian J.
Morsony\altaffilmark{2,3} and Mitchell C. Begelman\altaffilmark{2,4}}
\email{davide\_lazzati@ncsu.edu}
\shortauthors{Lazzati et al.}

\altaffiltext{1}{Department of Physics, NC State University, 2401
Stinson Drive, Raleigh, NC 27695-8202} 
\altaffiltext{2}{JILA, University
of Colorado, 440 UCB, Boulder, CO 80309-0440}
\altaffiltext{3}{Department of Astronomy, University of
Wisconsin-Madison, 5534 Sterling Hall, 475 N. Charter Street, Madison WI
53706-1582}
\altaffiltext{4}{University of Colorado, Department of Astrophysical and 
Planetary Sciences, 389 UCB, Boulder, CO 80309-0389}

\begin{abstract} 
We numerically analyze the evolution of a long-duration gamma-ray burst
jet as it leaves the progenitor star and propagates to the photospheric
radius, where radiation can be released. We find that the interaction of
the relativistic material with the progenitor star has influences well
beyond the stellar surface. Tangential collimation shocks are observed
throughout the jet evolution, out to about 100 stellar radii, which is
the whole range of our simulation. We find that the jet is internally
hot at the photospheric radius and we compute the photospheric emission.
The photosphere is a very efficient radiator, capable of converting more
than half of the total energy of the jet into radiation. We show that
bright photospheres are a common feature of jets born inside massive
progenitor stars and that this effect can explain the high radiative
efficiency observed in long-duration bursts. 
\end{abstract}


\section{Introduction}

Long-duration gamma-ray bursts (GRBs) are believed to arise when a very
massive star collapses to a black hole, powering a hyper-relativistic
outflow (Woosley 1993; Hjorth et al. 2003; Stanek et al. 2003; Malesani
et al. 2004; Woosley \& Bloom 2006). They are the brightest explosions
in the present day universe, releasing, in a matter of few tens of
seconds, a large amount of energy ($\sim10^{52}$~erg) in the form of a
hot, possibly magnetized, outflow. GRBs are able to radiate a large
fraction of their energy in the form of photons, with an efficiency
sometimes approaching 100\% (Zhang et al. 2007).

The commonly accepted model for the dissipation of the jet energy and
its conversion into radiation is the internal shock synchrotron model
(Rees \& Meszaros 1994). In this model,  differential velocities within
the outflow are dissipated through collisionless shock waves that
generate strong magnetic fields, accelerate a population of relativistic
electrons, and eventually release synchrotron radiation (e.g., Piran
1999; Lloyd \& Petrosian 2000). Internal shocks suffer however from poor
efficiency (Kobayashi et al. 1997; Lazzati et al. 1999; Spada et al.
2000), being capable of radiating only a few per cent of the total
energy in the outflow.

In their early evolution, the  ouflows of long-duration GRBs interact
strongly with the cold and dense material of their progenitor star
(MacFadyen \& Woosley 1999; Aloy et al. 2000; Zhang et al. 2003, 2004).
The interaction strongly modifies the dynamics of the jet while it is
still inside the star. Evaluating the effects of the jet-star
interaction at larger scales, when the jet is expanding in the
circum-stellar material, is of fundamental importance, since a change in
the jet dynamics can profoundly affect our understanding of the
dissipation and radiation processes at play in the prompt phase of
gamma-ray bursts (Rees \& Meszaros 2005; Ghisellini et al. 2007).

Here we present the results of a hydrodynamic simulation that follows a
gamma-ray burst jet out to 65 stellar radii, where the jet material
becomes transparent to radiation. Our calculation shows that the
interaction of the jet material with the star profoundly affects the
subsequent evolution of the jet, generating continuous dissipation
through transverse collimation shocks.

This letter is organized as follows: in \S~2 we present the numerical
model adopted for the jet and the progenitor star, in \S~3 we present
our result and compute light curves, and in \S~4 we discuss our
findings.

\section{Numerical model}

We considered a 16 solar-mass Wolf-Rayet progenitor star, evolved to
pre-explosion (Model 16TI, Woosley \& Heger 2006). The jet was
introduced as a boundary condition at a distance of $10^9$~cm from the
center of the star, with a total luminosity
$L_{\rm{jet}}=5.33\times10^{50}$~erg/s, an initial opening angle
$\theta_0=10^\circ$, an initial Lorentz factor $\Gamma_0=5$, and a ratio
of internal over rest mass energy $\eta_0=80$, allowing for a maximum
Lorentz factor of $\Gamma_\infty=\Gamma_0\eta_0=400$, reachable in case
of complete, non-dissipative acceleration. The jet evolution was
computed with the special-relativistic hydrodynamic code FLASH (Fryxell
et al. 2000).

\begin{figure}
\plotone{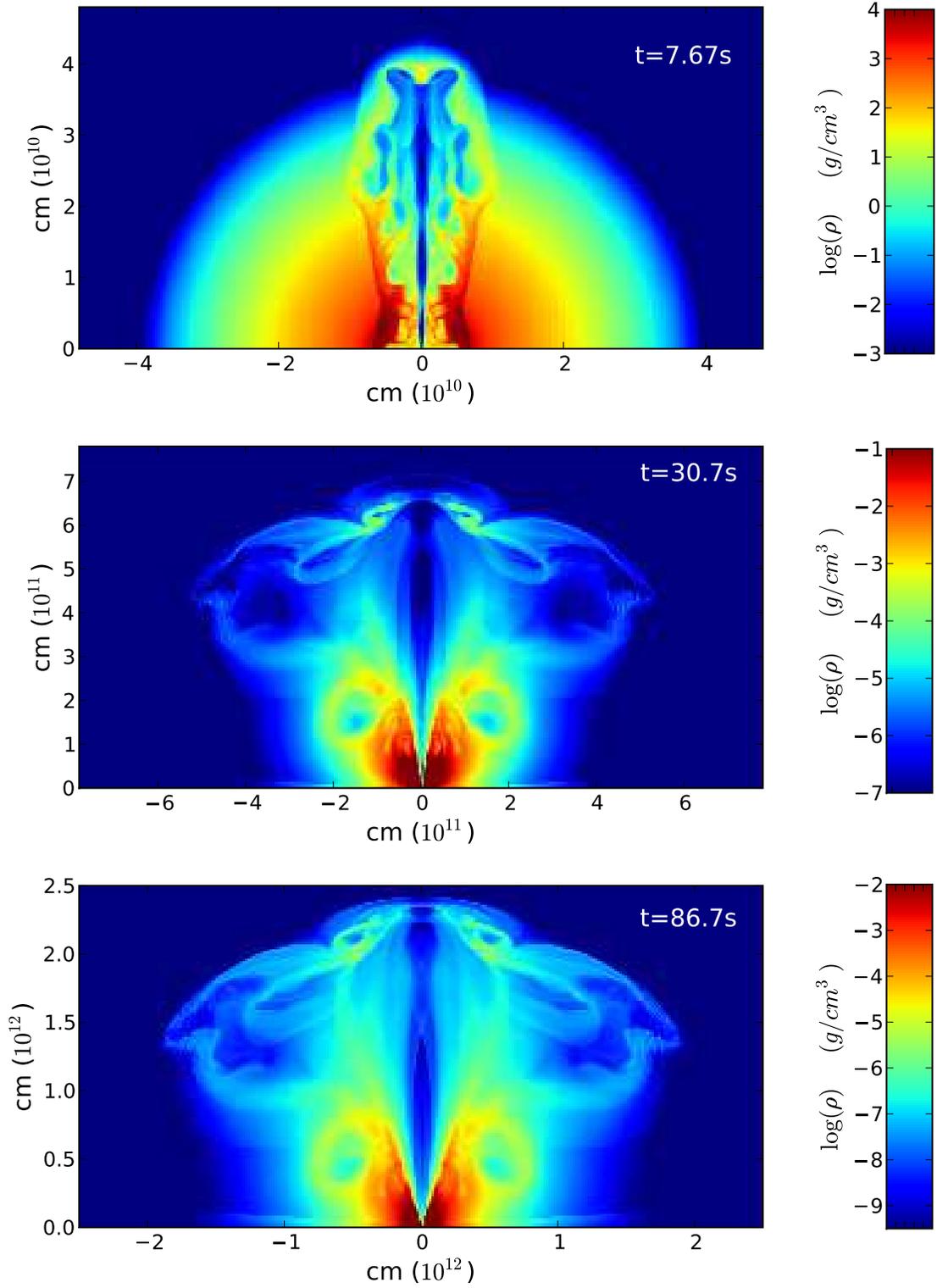}
\caption{{Stills from our simulation of a GRB jet expanding through a 16
solar-mass Wolf-Rayet progenitor star. The jet initially bores a hole in
the stellar material (upper panel) and subsequently evolves out to 20
stellar radii (middle panel) and eventually to 60 stellar radii (lower
panel). The images show logarithmic comoving density, with the density
color scale specified in the right column. The density of the jet in
the last panel is sufficiently low for photons to escape to infinity,
and the photospheric component of the GRB light curve can be directly
computed.} 
\label{fig:frames}} 
\end{figure}

Figure~\ref{fig:frames} shows density images from three frames of the
simulation at $t=7.7$~s, $t=30.7$~s, and $t=80.7$~s. The upper panel
shows the collimating power of the stellar material. The jet that was
injected with an opening angle of 10 degrees is squeezed into a 2-degree
opening angle, with the typical sausage shape due to the presence of
strong tangential collimation shocks (Morsony et al. 2007). This
simulation extends into a box ten times bigger than any previously
performed in a single run and was run for 150 seconds. The maximum
spatial resolution along the jet axis was $4\times10^{6}$~cm, while the
maximum spatial resolution in the outer jet was $2.5\times10^8$~cm. In
the central and bottom panels it is possible to see that the jet remains
affected by the shocks out to many stellar radii (20 in the central
panel and about 60 in the bottom panel).

\section{Photospheric emission and light curves}

\begin{figure}
\plotone{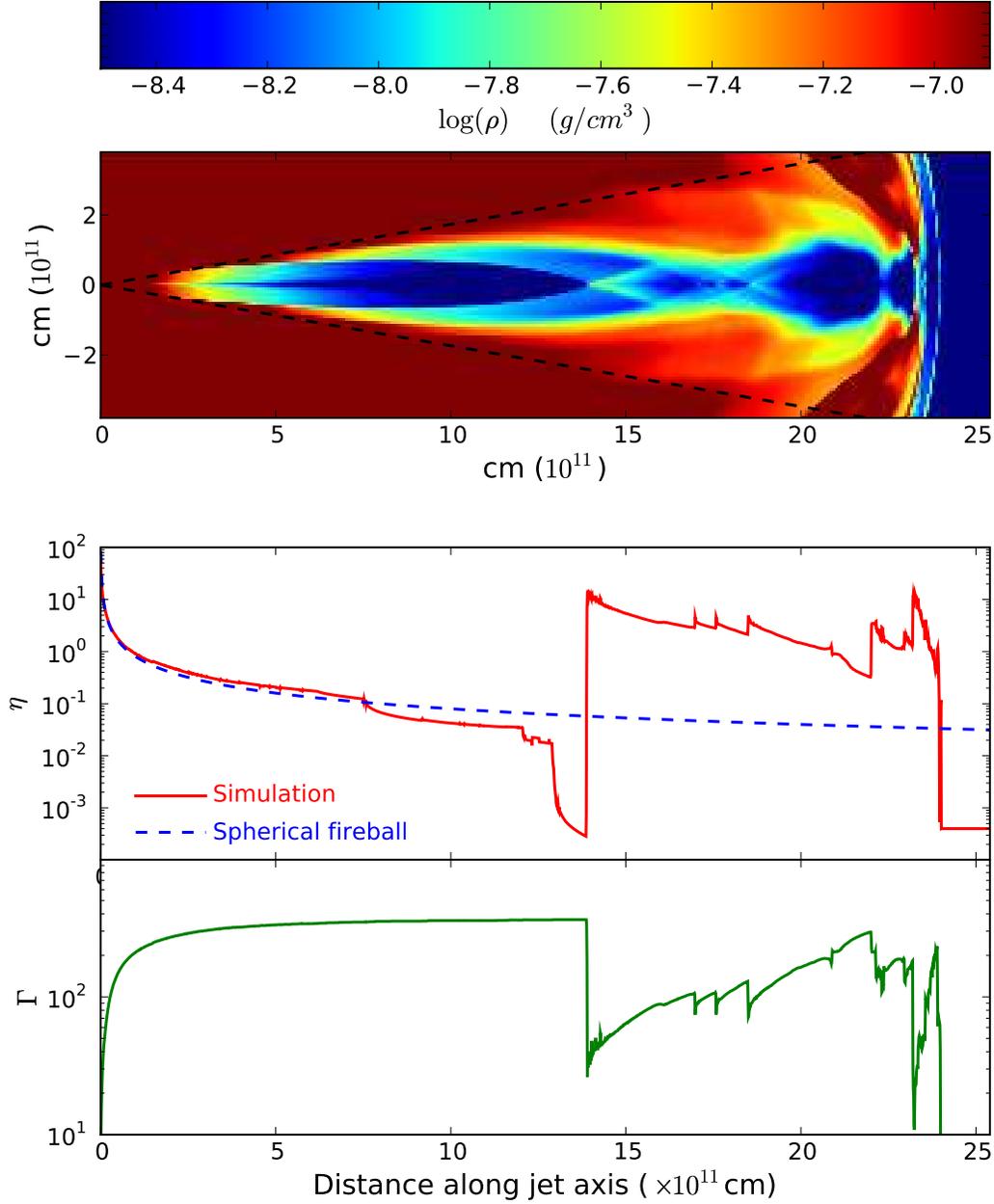}
\caption{{Enhanced dissipation due to shock activity at large radii.
This figure is a zoom of the jet region of the lower panel of
Fig.~\ref{fig:frames}. The figure shows logarithmic comoving density
with a color scale enhanced to highlight the shock structure that
forms along the jet. Dashed lines show the projected opening angle of
the jet for a free expansion, in which the jet keeps the opening angle
with which it was injected ($\theta_0=10^\circ$). The outer part of the
jet, on the right in the image, shows a structure of mainly parallel
shocks, in which the normal to the shock front is parallel to the jet
axis and to the flow velocity. The intermediate portion of the jet shows
instead a structure dominated by tangential collimation shocks. Finally,
the inner part of the jet, on the left in the image, shows a more
relaxed structure, due to the progressive disappearance of the stellar
progenitor's collimating power. The middle panel shows the ratio of the
internal energy density to the rest mass energy density along the jet
axis. The blue line shows the prediction for a freely expanding fireball
(a conical jet), while the red line shows the results of our simulation.
The effect of shocks and a rarefaction wave (at about
$1.4\times10^{12}$~cm) are clearly identifiable. The bottom panel shows
the Lorentz factor along the jet spine.} 
\label{fig:shocks}}
\end{figure}

A zoom around the jet region in the last panel of Fig.~\ref{fig:frames}
shows the effect of the interaction of the outflow with the progenitor
material more quantitatively (Fig.~\ref{fig:shocks}).  The head of the
jet shows shocks whose normal is parallel to the jet axis and propagate
along the jet (similar to internal shocks or to the scenario of reborn
fireballs, Ghisellini et al. 2007),  while the part of the jet behind
the head is characterized by tangential shocks that propagate from the
side of the jet to its axis and vice versa. The middle panel shows the
ratio of the internal energy to the rest mass energy. The dashed blue
line shows the theoretical prediction for a freely expanding outflow,
neglecting the effect of the interaction with the progenitor material.
In that case, $\eta=\eta_0 (R_0/R)$, where $\eta_0$ is the ratio of
internal to rest energy when the jet is released at a radius $R_0$. The
red line shows the result of our simulation along the jet axis. Note
that the horizontal scales are equal among panels and the one to one
correspondence between the shocks and the increases in internal energy.
The internal energy dominates over the rest mass energy much farther out
than the simple theory predicts, being $\sim 400$ times larger in the
region between $R=1.5\times10^{12}$ and $R=2.5\times10^{12}$~cm. The
bottom panel of Fig.~\ref{fig:shocks} shows the Lorentz factor along the
jet spine.

The opacity for Thompson scattering of a relativistically moving medium
is (Abramowicz et al. 1991): 
\begin{equation} 
\tau(R,t)=\sigma_T\int_R^\infty \Gamma
\left(1-\beta\cos\theta_v\right)n^\prime dr 
\end{equation}
where the comoving density $n^\prime$, the Lorentz factor $\Gamma$ and
the angle between the line of sight and the velocity of the fluid
($\theta_v$) are evaluated at a laboratory time
$t_{\rm{lab}}=t_{\rm{lab,0}}+R/c$, where $t_{\rm{lab,0}}$ is the
laboratory time at which the photon is released at a radius $R$. 
The photosphere is defined as the radius $R_{\rm{phot}}$ for which 
$\tau(R,t)=1$.

\begin{figure} 
\plotone{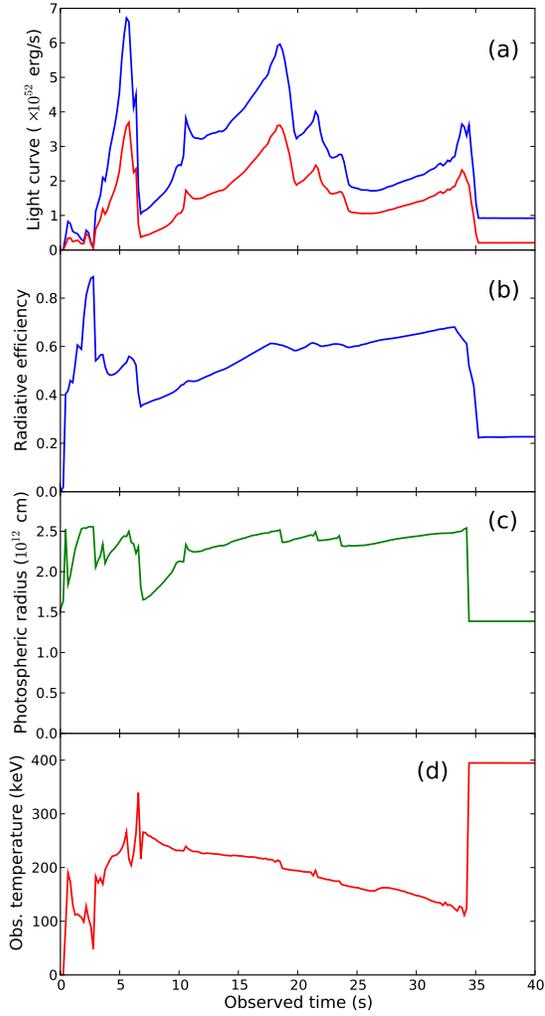} 
\caption{{Photospheric light curve and properties for an on-axis
observer at infinity. Panel (a) shows the total light curve (blue line)
and the photospheric component (red line) for an observer lying along
the jet axis. Panel (b) shows the photospheric fraction, i.e., the
fraction of the total energy that is released at the photosphere. This
represents a lower limit to the radiative efficiency since additional
dissipation and radiation of non-thermal photons can take place beyond
the photosphere. Panel (c) shows the radius of the photosphere that
contributes photons at time t along the jet axis. Finally, panel (d)
shows the observed temperature of the photospheric component in keV. The
predicted temperature is strikingly similar to the typical peak
frequency of observed GRBs. }
\label{fig:lcur}} 
\end{figure}

Knowing the photospheric radius as a function of time and angle we
computed the light curve by integrating a thermal spectrum over the
angular direction at the comoving temperature boosted by relativistic
effects:
\begin{equation}
L=\frac{ac}{2}\int_0^{\pi/2}
\frac{T^{\prime4}R^2\sin(\theta)d\theta}{\left(1-\beta\cos\theta_v\right)^2}
\end{equation}
where $a=7.56\times10^{-15}$~erg~cm$^{-3}$~K$^{-4}$ is the radiation
constant and $T^\prime$ is the comoving radiation temperature, which is
derived from the pressure as $T^\prime=(3p/a)^{1/4}$. Results are shown in 
Fig.~\ref{fig:lcur}. The photospheric light curve (panel a) shows an intensity
and a temporal evolution fairly similar to observed GRBs. 

It is worth noting here that the jet material is fully ionized and
opacity is mainly provided by Thompson scattering, with a minor
contribution from free-free processes. In such conditions, the spectrum
is slightly broader than a black body (Goodman 1986), but has the same
peak frequency and fairly similar asymptotic behavior. In addition, a
population of non-thermal particles could be present due to the shocks
that keep the photosphere hot. Such non-thermal electrons would create a
non-thermal tail at high enrgies due to inverse Compton scattering.
However, if the high energy tail extend to comoving energies
$h\nu^\prime>511$~keV, photon-photon interaction may become relevant as
a source of opacity for high energy photons as well as a source of
electron-positron pairs that could alter the location of the photosphere
(e.g., Meszaros et al. 2002). The details of the opacity calculation
depend on the assumed non-thermal tail of the spectrum. In this letter,
we concentrate on the thermal emission from the photosphere and
therefore the pairs are not an issue.  The explanation of GRB spectra
extending to energies larger than $h\nu\sim50 (\Gamma/100)$~MeV require
a more detailed calculation and assumptions on the spectrum of
non-thermal particles that go beyond the scope of this letter.

The kinetic energy remaining in the jet beyond the photosphere was used
to evaluate the minimum radiative efficiency. It is about 50 per cent,
with an average of 56 per cent in the first 40 seconds of the light
curve (see Fig.~\ref{fig:lcur}, panel b). Figure~\ref{fig:lcur}, panel
(c), shows the photospheric radius along the jet axis as a function of
time, while panel (d) shows the observed peak of the photospheric
emission in keV. The peak of a few hundred keV seen in the simulation is
typical for long-duration GRBs (Kaneko et al. 2006). Note that at times
longer than $\sim35$~s the jet enters the ``unshocked jet phase", as
discussed in Morsny et al. (2007). In this phase, the photospheric
efficiency drops since tangential shock are absent in the ``unshocked"
phase.

\section{Discussion}

Our simulation shows that the interaction of the relativistic outflow
with the progenitor material in long duration GRBs has effects that are
important well after the jet has left the star. The jet does not cool
and has a very bright photosphere yielding a high radiative efficiency
in the prompt phase. In addition, the photospheric component is
quasi-thermal (Rees \& Meszaros 2005; Thompson et al. 2007; Pe'er 2008)
offering an explanation for the hard X-ray spectra in the early phase of
gamma-ray bursts, which is inconsistent with synchrotron (Preece et al.
1998, Ghisellini et al. 2000). Thermal components have been detected
recently in BATSE spectra (Ghirlanda et al. 2003, 2007; Ryde et al.
2005; Ryde \& Pe'er 2008) and the Fermi gamma-ray telescope will soon
give us a wideband view of the bursts' prompt emission. Non-thermal
components can be added to the spectrum as a result of the reprocessing
of photospheric emission and of additional dissipation due to the large
inhomogeneities in the jet Lorentz factor (see bottom panel of Fig. 2).
Additional work is however required to assess the capability of this
model to explain the very high-energy photons observed, e.g., in
GRB~080916C (Abdo et al. 2009). Unless an extremely large value of the
Lorentz factor of the fireball is assumed ($\Gamma>10^4$ to explain
photons of 10 GeV), photons of GeV energy need to be released at a
radius of at least $\sim10^{15}$~cm (Zhang \& Pe'er 2009). Photospheres
cannot be located at such large radii, even if the opacity is increased
by electron-positron pair production. If the keV to MeV photons of
GRB~080916C are from the photosphere, as we argue, a second emission
mechanism is required for the GeV component. Kumar \& Barniol Duran
(2009) argue that the $>100$~MeV photons observed in GRB~080916C must be
produced in the external shock due to their temporal evolution that is
different from the one of the keV to MeV photons. In such a scenario,
the keV to MeV radiation of GRB~080916C could be photospheric. It is
also possible that photons of the photospheric emission are inverse
Compton scattered by hot electrons at large radii (heated, e.g., in an
internal shock) to produce a high energy tail in the GeV band.

\begin{figure}
\plotone{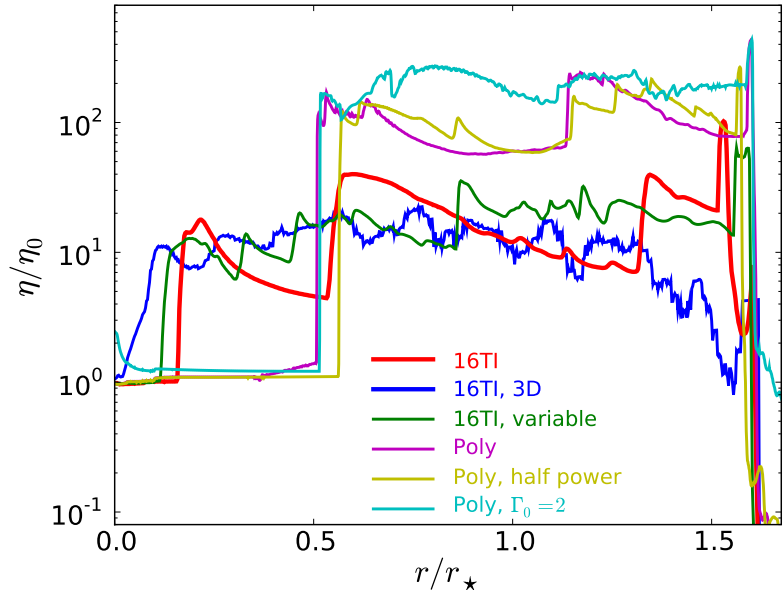}
\caption{{Comparison between the excess internal energy ratio of
different jet/progenitor models. The vertical axis shows the ratio
between the actual internal to rest mass energy ($\eta$) over the
internal to rest mass energy ratio in a freely expanding jet ($\eta_0$).
The horizontal axis shows the distance along the jet in units of the
progenitor star radius. The reference model discussed in the letter is
shown with the thick red line and labelled "16TI". Model "16TI, 3D"
shows an identical physical setup, but computed in three dimensions.
Model "16TI, variable" shows a jet with the same average luminosity, but
with short time-scale variations, expanding in the reference progenitor.
Model "poly" shows the reference jet expanding in a bigger star
($r_\star=10^{11}$~cm) with a polytropic density profile and a total
mass of 15 $M_\odot$. Model "poly, half power" shows a jet with half the
luminosity of the reference jet expanding in the poly progenitor. Model
"poly, $\Gamma_0=2$" shows a jet with an initial Lorentz factor of 2
(compared to $\Gamma_0=5$ of the reference jet) propagating through the
poly progenitor.}
\label{fig:other}} 
\end{figure}

Our computation has two main limitations. On the one hand, it represents
one single case of a jet, with given characteristics, interacting with a
given star. It is not straightforward to predict how different would be
the photospheric component for a different jet with a different
progenitor star. Based on the result of simulations of smaller domain
(Fig.~\ref{fig:other}) we can speculate that bigger stars will produce a
stronger photospheric component and that the diversity in progenitor
population would explain the range of thermal to non-thermal ratios and
efficiencies observed in gamma-ray bursts. Higher luminosity jets are
also expected to produce brighter photospheres, since they produce
thicker cocoons (Aloy et al. 2000). Mizuta \& Aloy (2008) performed a
systematic study of jets into a diverse set of progenitor stars, finding
that the early jet evolution is largely affected by the structure of the
stellar progenitor. On the other hand, the simulation does not have a
magnetic component, but is still debated whether the launching and
collimation of the jet are due to magnetic effects (e.g., McKinney 2006)
or not (see, e.g., Aloy \& Obergaulinger 2007). Despite these
limitations, our result opens a new window for the understanding of
radiation from GRB jets.

\acknowledgements We thank the anonymous referee for a careful review
that improved the clarity and content of this letter. We would like to
thank Chris Fryer, Ehud Nakar, Asaf Pe'er, Daniel Proga, Felix Ryde, Jay
Salmonson, and Bing Zhang for insightful discussions. The software used
in this work was in part developed by the DOE-supported ASC / Alliance
Center for Astrophysical Thermonuclear Flashes at the University of
Chicago. This work was supported in part by NASA ATP grant NNG06GI06G
and Swift GI program NNX06AB69G and NNX08BA92G. We thank NCSA and the
NASA NAS for the generous allocations of computing time.

\newpage

\end{document}